\documentclass[sigconf,10pt]{acmart}
\setcopyright{none}
\renewcommand\footnotetextcopyrightpermission[1]{}

\usepackage{paralist}
\usepackage{algorithm}
\usepackage{algpseudocode}
\usepackage{tcolorbox}
\usepackage{multirow}
\newtcolorbox[auto counter, number within=section]{rqbox}[1][]{colback=gray!9!white,           colframe=black, 
    boxrule=0.3mm, 
    arc=3mm, 
    auto outer arc,
    boxsep=0mm,
    #1
}

\newcommand*\circled[1]{%
  \tikz[baseline=(char.base)]{
    \node[shape=circle, draw, inner sep=0.8pt, line width=0.3pt, font=\small] (char) {#1};
  }%
}

\usepackage{enumitem}

\AtBeginDocument{%
  }

\begin{document}

\author{Chanwoo Cho}
\email{chanwoo_cho@korea.ac.kr}
\affiliation{%
  \institution{Korea University}
  \city{Seoul}
  \country{Republic of Korea}
}
\author{Wooseok Kim} 
\email{ws8462@korea.ac.kr}
\affiliation{%
  \institution{Korea University}
  \city{Seoul}
  \country{Republic of Korea}
}
\author{Yonglak Son}
\email{yonglak_son@korea.ac.kr}
\affiliation{%
  \institution{Korea University}
  \city{Seoul}
  \country{Republic of Korea}
}
\author{Young Seo Lee}
\authornote{Co-corresponding author.}
\email{yslee@ssu.ac.kr}
\affiliation{%
  \institution{Soongsil University}
  \city{Seoul}
  \country{Republic of Korea}
}

\author{Young Geun Kim}
\authornotemark[1]
\email{younggeun_kim@korea.ac.kr}
\affiliation{%
  \institution{Korea University}
  \city{Seoul}
  \country{Republic of Korea}
}



\begin{abstract}
Large language models (LLMs) are widely used in intelligent services due to their remarkable capability in generative tasks. Typically, LLM-based services process the inference requests of the users in a centralized data center. Unfortunately, such centralized execution has limitations for end-users, such as increased response latency with communication overhead and privacy leakage risk. To alleviate the aforementioned limitations, there have been increasing pushes to execute LLM inference locally on user-end devices. However, the limited resources of a single edge device impose restrictions on achievable accuracy of LLMs. To overcome the issue, we first propose to leverage multiple user-end devices available at the edge for LLM inference, enabling the execution of larger models. Specifically, we propose Voltron, a novel on-device LLM inference framework that elastically utilizes multiple user-end devices for LLM inference execution while adapting to diverse real-world edge environments. In our evaluation, Voltron achieves up to 16.5\% higher accuracy than state-of-the-art LLMs that can be executed on a single edge device, satisfying user QoS requirements.
\end{abstract}
\title{Voltron: Enabling Elastic Multi-Device Execution of LLM Inference for Empowered Edge Intelligence}
\maketitle
\pagestyle{plain}

\section{Introduction}


Large language models (LLMs) have shown a remarkable capability in generative tasks such as question answering, text generation, and summarization. Many intelligent services are utilizing the capability of LLMs as a key functionality. For example, personal assistants (e.g., Siri~\cite{AppleSiri} and Alexa~\cite{amazonAlexa}) use LLMs to improve their conversational abilities, AI chatbots (e.g., ChatGPT~\cite{ChatGPT} and Gemini~\cite{Gemini}) provide information through question-and-answer interactions using LLMs, and personal agents (e.g., OpenClaw~\cite{OpenClaw}, ManusAI~\cite{ManusAI}) rely on LLMs to autonomously plan and execute complex user tasks.

Typically, LLM-based services batch inference requests from tens of millions of users in a centralized data center~\cite{slora, vllm, dlora}. To accelerate the large volume of inference requests, state-of-the-art inference serving frameworks such as vLLM~\cite{vllm} are used to maximize the size of batched LLM inference requests. However, such centralized execution has several limitations for end-users. First, the response latency can be prolonged by communication overhead depending on wireless network conditions and congestion in core network, which adversely impacts a user QoS --- the communication overhead can further be increased depending on the subscription option of the service~\cite{GPTPricing, GeminiPricing}. Second, the inference requests may contain user private data, posing a security concern~\cite{GPTPrivacy}.


To alleviate the aforementioned limitations, there have been increasing pushes to execute LLM inference locally on user-end devices by leveraging the available computational resources recently featured in mobile SoCs --- on-device execution of LLM inference is expected to open up new personalized applications, such as smart home agent~\cite{amazonAlexa}, personal AI robot~\cite{neohomerobot}, etc. However, it is still not feasible to execute large models whose required memory is beyond the device memory capacity. To enable on-device LLM, many researchers have focused on designing smaller LLMs (sLLMs) (e.g., MobileLLM~\cite{mobilellm} and TinyLlama~\cite{tinyllama}) or applying model compression techniques such as quantization~\cite{AWQ, SmoothQuant, GPTQ} and pruning~\cite{LLMPruner, SixteenHead}. However, such techniques inevitably cause accuracy drop due to information loss.


To break through the memory constraint, we first propose to leverage multiple nearby edge devices for LLM inference execution. According to~\cite{CiscoAnnualInternetReport}, mobile users typically use multiple devices (3.6, on average) at the same time, including smartphones, tablets, and smartwatches. Furthermore, there can be various IoT (Internet of Things) devices, such as smart speakers and smart kitchen appliances, available at the smart home. By exploiting such available multiple devices simultaneously, it is possible to execute larger LLMs achieving higher inference accuracy. 


To enable multi-device execution of LLM inference, distributed inference methods, such as tensor parallelism~\cite{megatronlm}, can be used. However, our characterization reveals that adopting the distributed inference methods in edge environment poses two key challenges as follows:

\begin{itemize}[label=--, leftmargin=*, nosep]
\item\textbf{Heterogeneity:} Edge environments inherently exhibit multiple forms of heterogeneity. Devices in edge clusters often have highly heterogeneous computational and memory capabilities, making the execution characteristics of the distributed inference methods vary across the devices. Furthermore, the characteristics are also affected by LLM architectures, and mixed precision configurations that are natural in resource-constrained edge environment. These factors make it challenging to apply a single distributed inference method as a one-size-fits-all solution in edge.

\item\textbf{Runtime variance:} Edge LLM inference execution is stochastic by nature. The computational characteristics of LLM inference can vary across inference phases and conversations with various input/output length. Wireless network variability also significantly affects the communication overhead of distributed inference, due to signal strength fluctuations caused by user movement. Such runtime variance makes it challenging to efficiently employ appropriate distributed inference methods in continuously changing execution environments.

\end{itemize}


To tackle the aforementioned challenges, we propose Voltron, a novel on-multi-device LLM inference execution framework that enables elastic execution of LLM across edge devices. Whenever a user types a new prompt for LLM inference, Voltron observes the current execution environment and quickly determines an appropriate execution strategy for that environment. During LLM inference, Voltron adapts to changing environments by adjusting the execution strategy to satisfy QoS requirements while maximizing accuracy.

The key contributions of this work include:
\begin{itemize}[label=--, leftmargin=*, nosep]

\item We present an in-depth performance and accuracy characterization of multi-device LLM inference at the edge. The results show that performance and accuracy of multi-device LLM inference can significantly vary under multiple forms of heterogeneity and runtime variance (Section~\ref{Section3}).

\item We propose Voltron, a novel LLM inference execution framework that elastically utilizes multiple user-end devices available at the edge. Voltron efficiently determines the appropriate execution strategy for the current environment and continuously adjusts it to adapt to changing environments, maximizing accuracy while satisfying QoS requirements (Section~\ref{Section4}).

\item We implement Voltron\footnote{We plan to open-source Voltron upon acceptance to expedite the adoption of multi-device LLM execution for empowered edge intelligence.} across various combinations of edge clusters and models. In our evaluation, Voltron achieves 10.2\% higher accuracy on average (up to 16.5\% higher accuracy) than state-of-the-art LLMs that can be executed on a single device, while satisfying the QoS requirements even in the presence of runtime variance (Section~\ref{Section5}).
\end{itemize}

\section{Background}
\label{Section2}

\begin{figure}[t!]
\centering
\includegraphics[width=\linewidth]{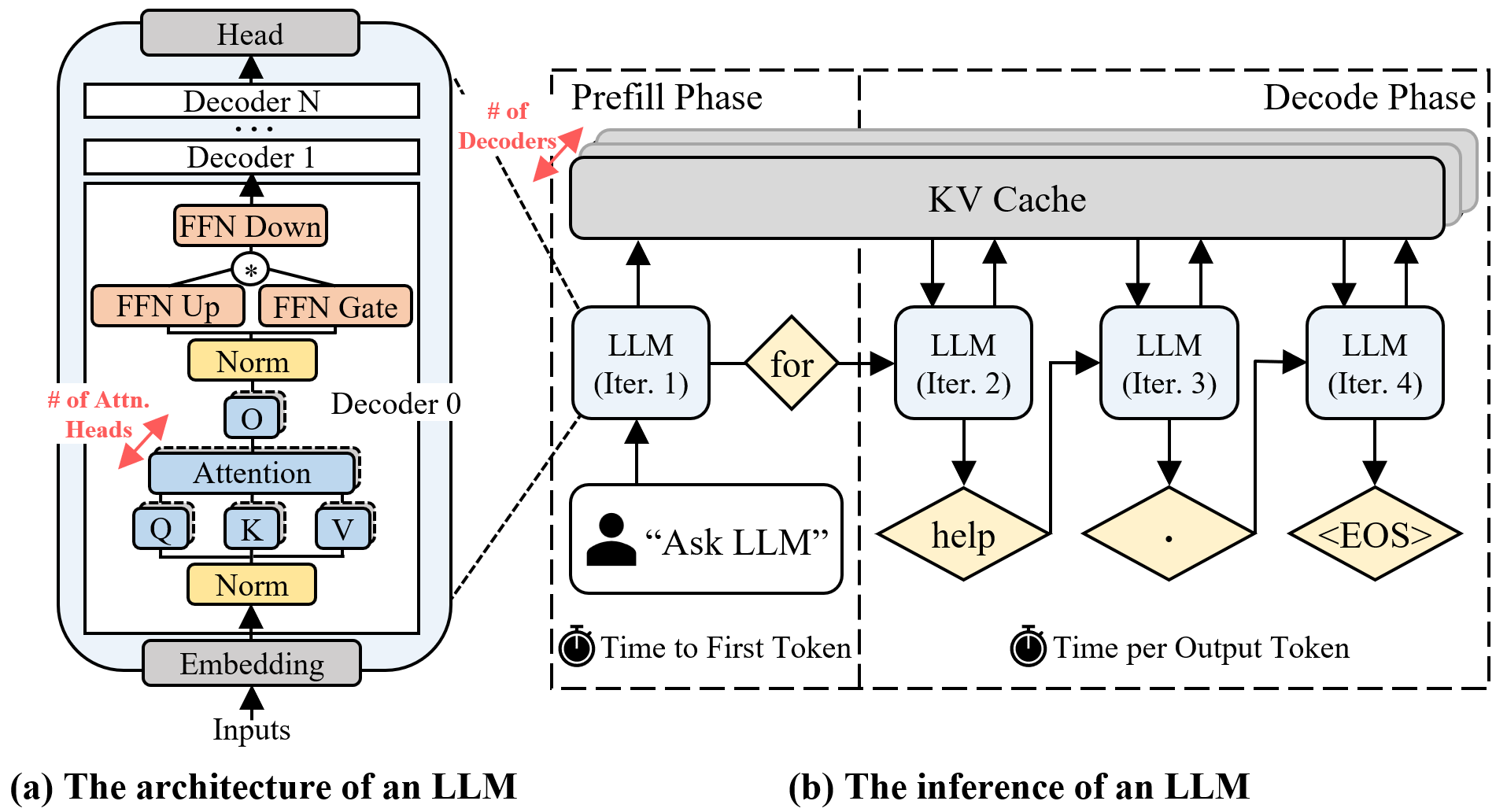}
\caption{LLM architecture and inference process.}
\label{fig:Background1}
\end{figure}

\subsection{Large Language Model}

\textbf{LLM Architecture: }An LLM generally has an architecture where identically structured decoders (depicted in Fig.~\ref{fig:Background1}(a)) are stacked. When a user enters an input prompt, the prompt is converted into LLM inputs (i.e., tokens) through the tokenizer. The tokens are transformed into high-dimensional vectors (i.e., input embeddings), and then pass through the attention and feed-forward layers~\cite{AttentionIsAllYouNeed}. 

In the attention layer (ATTN) (highlighted with blue color in Fig.~\ref{fig:Background1}(a)), each attention head independently computes an attention score with scaled dot-product operations between query (Q), key (K), and value (V) matrices. By using multiple attention heads, the attention mechanism projects Q, K, and V into different subspaces, which allows the LLM to focus on different parts of an input sequence simultaneously. The attention outputs are then passed through an output linear (O) and subsequently processed by the feed-forward network (FFN), which projects them into higher-dimensional feature spaces. The feed-forward network further applies nonlinear transformations that refine each token representation and improve the model reasoning capability.

Recent LLM architectures are also employing several variants of ATTN and FFN for efficient parameter scaling~\cite{ainslie2023gqatraininggeneralizedmultiquery, beltagy2020longformerlongdocumenttransformer, shazeer2017outrageouslylargeneuralnetworks}. For example, group query attention (GQA) reduces computational load of multi-head attention (MHA) by sharing K and V projections across multiple query heads, while sliding window attention (SWA) reduces the load by restricting attention to a fixed-size local window. For FFN, mixture of experts (MoE) is widely used to reduce the load by activating only a subset of expert networks for each token.


\noindent \textbf{LLM Inference: } Fig.~\ref{fig:Background1}(b) shows the LLM inference process. The inference of LLM is divided into two phases: the prefill and decode phases. In the prefill phase, the LLM processes the tokens of a user's input prompt (e.g., "Ask LLM" in Fig.~\ref{fig:Background1}(b)) in parallel and produces the first output token (e.g., "for" in Fig.~\ref{fig:Background1}(b)). The LLM saves the calculated K and V tensors of the input tokens in the KV cache. The latency to complete the prefill phase is called time-to-first-token (TTFT). In the decode phase, the LLM generates one output token per iteration in an autoregressive manner. A new token (e.g., "help" in Fig.~\ref{fig:Background1}(b)) is jointly predicted based on the tensors of the generated token (e.g., "for" in Fig.~\ref{fig:Background1}(b)) and those stored in the KV cache (e.g., "Ask LLM" in Fig.~\ref{fig:Background1}(b)). Meanwhile, K and V tensors of the most recently generated token (e.g., "for" in Fig.~\ref{fig:Background1}(b)) are accumulated in the KV cache to be used for the next token generation. This process repeats until the “$<$EOS$>$” token is generated as the output. The latency of the decode phase is called time-per-output-token (TPOT).

\subsection{On-Device LLM}
LLM inference has traditionally been executed in centralized data centers due to their large memory and computation requirements. Recently, however, there is a growing demand for executing LLM inference locally on edge devices, driven by the emergence of LLM applications that require long contexts containing privacy sensitive user information, such as private personal agents~\cite{zhang2025personaagentlargelanguagemodel} and autonomous household robots~\cite{neohomerobot}. To enable LLM inference on a single edge device with limited resources, prior works have focused on reducing the computational and memory demands of LLM inference through algorithm-side techniques such as designing sLLMs~\cite{mobilellm, tinyllama}, quantization~\cite{AWQ, SmoothQuant}, and pruning~\cite{LLMPruner, SixteenHead}. However, such approaches often result in low accuracy, along with long TTFT and TPOT.

\begin{figure}[t!]
\centering
\includegraphics[width=\linewidth]{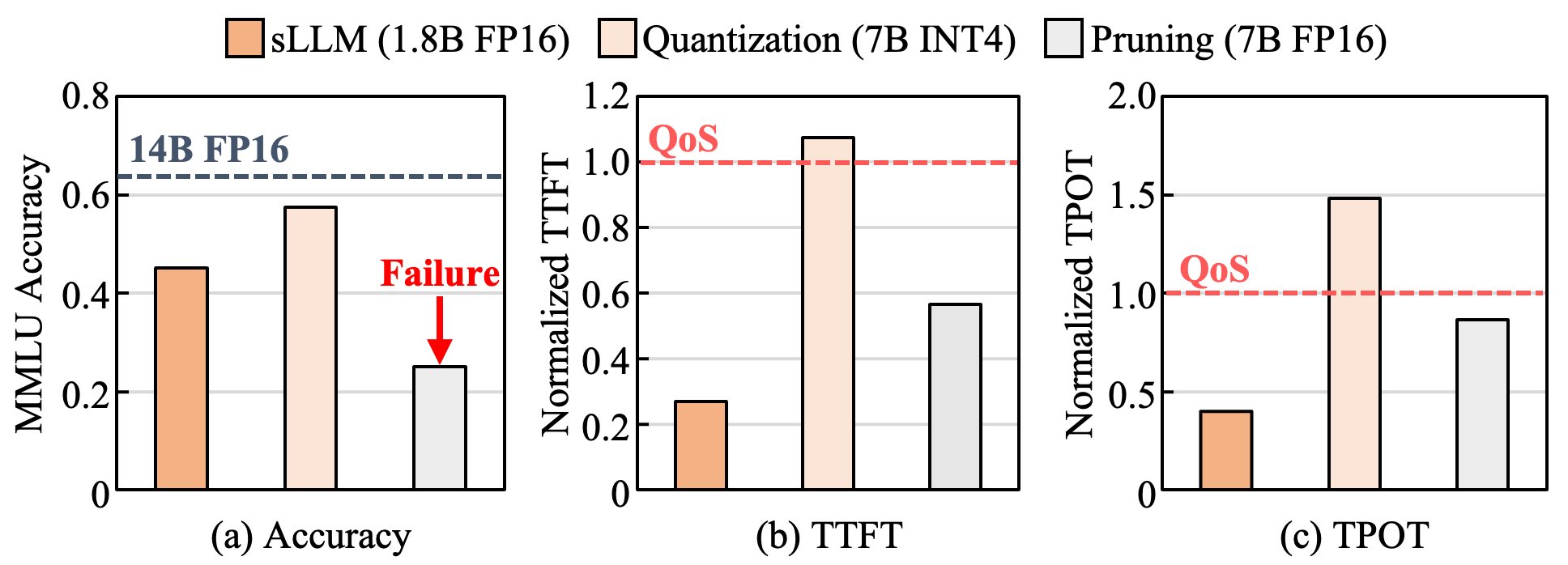}
\caption{Accuracy and performance of on-device LLM techniques. Existing techniques fail to satisfy accuracy and QoS requirements.}
\label{fig:Background2}
\end{figure}

Fig.~\ref{fig:Background2} shows the accuracy, TTFT, and TPOT of LLMs achieved through representative algorithm-side techniques --- TTFT and TPOT are normalized to respective QoS constraints. sLLM is a model designed with a small number of parameters to execute effectively in resource-constrained environments. Though sLLM may meet the QoS constraints (i.e. TTFT of 10s~\cite{sageserve} and TPOT of 400ms\footnote{Prior studies show that users expect mobile devices to respond in several hundred milliseconds, and their satisfaction quickly drops as latency grows beyond around 400ms~\cite{10.1007/978-3-030-23535-2_30}.}) due to reduced computation load, it exhibits much lower accuracy compared to the LLM (i.e., 14B LLM)~\cite{chinchilla}. On the other hand, quantization reduces the numerical precision of parameters from high precision (e.g., FP16) to lower bit representations (e.g., INT4), thereby reducing both the computation load and memory footprint. The quantized LLM, however, exhibits lower accuracy compared to the non-quantized LLM (i.e., 14B LLM), due to the fine-grained information loss~\cite{kuzmin2024pruningvsquantizationbetter}. It also fails to meet the QoS constraints since the number of parameters remains unchanged. Pruning reduces memory usage and computational load by removing a subset of weights (e.g., attention heads or layers). Although the pruned LLM satisfies the QoS constraints, it shows the lowest accuracy due to significant information loss, as the pruned heads or layers may contain important information. These results demonstrate that, given the limited resources of a single edge device, it is challenging to provide satisfactory accuracy while meeting the QoS requirements solely with the algorithm-side techniques.
\section{LLM Execution on Multi-edge Devices}
\label{Section3}

\subsection{Opportunities and Benefits}
\label{Section3.1}
To break through the limitations of single-device LLM inference execution, we first propose to leverage multiple nearby edge devices for LLM inference execution. According to~\cite{CiscoAnnualInternetReport}, mobile users typically interact with multiple devices (3.6, on average)---such as smartphones, tablets, and smartwatches---at the same time. The rapid growth of IoT technologies has also resulted in the spread of numerous smart home devices, including TVs, speakers, and sensors. Such devices are typically co-located and connected through low-latency wireless network, which enables fast data exchange across the devices~\cite{FluidXP}. The multiple devices can thus effectively form a local compute cluster. Along with the advancements of distributed inference execution methods, such as model parallelism (MP) and tensor parallelism (TP), the multi-device cluster opens new opportunities to cooperatively execute larger LLM models.

The multi-device execution offers several benefits for on-device LLM inference. By aggregating the memory and compute resources across devices, it becomes feasible to execute larger LLMs and improve inference accuracy compared to single-device execution. Moreover, the aggregated memory resources also enable long-context applications (e.g., personalized LLM agents~\cite{zhang2025personaagentlargelanguagemodel}) to be executed locally at the edge by accommodating larger KV caches. Executing LLM locally also allows users to include richer private information into the context without incurring privacy-leakage risks, enabling more personalized responses compared to server-side LLMs. These advantages highlight the strong potential of multi-device execution for enabling scalable and privacy-preserving LLM inference on edge devices.

\subsection{Challenges}
\label{Section3.2}

To enable LLM inference across multiple edge devices, distributed inference methods such as MP and TP can be used. However, naively adopting the distributed inference methods can result in inefficient execution of LLM inference, due to their computation-communication characteristics (Section~\ref{Section3.2.1}), device, model, and precision heterogeneity (Section~\ref{Section3.2.2}) and runtime variance (Section~\ref{Section3.2.3}). In this section, we characterize and quantify the impact of each factor on multi-device LLM execution\footnote{Note TTFT/TPOT are normalized to respective QoS constraints throughout the characterization results.}.

\begin{figure}[t!]
\centering
\includegraphics[width=\linewidth]{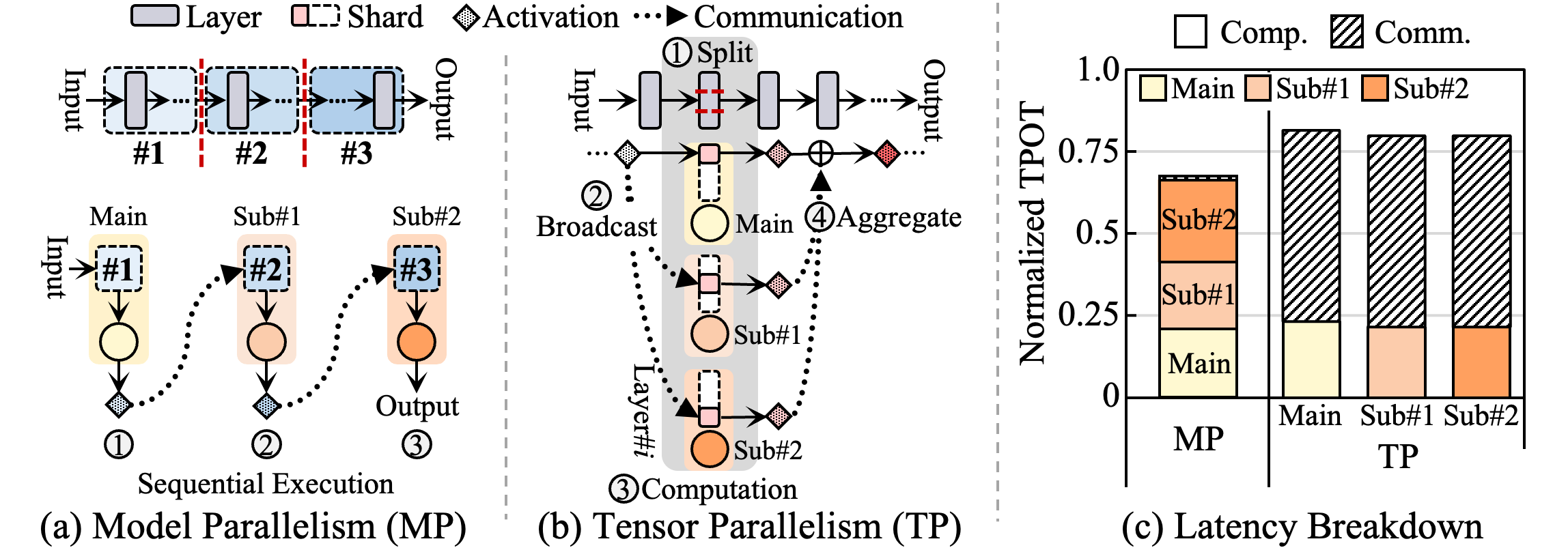}
\caption{Parallelism methods for multi-device LLM inference execution and TPOT breakdown of multi-device LLM inference under the homogeneous cluster.}
\label{fig:Motivation1}
\end{figure}

\subsubsection{Characteristics of Distributed Inference Methods}
~\label{Section3.2.1}

In this section, we characterize two representative distributed inference methods, MP and TP, on edge execution environment. These methods collaboratively execute LLM inference on the multiple edge devices, while sharing the intermediate data (e.g., output activations) via wireless network (e.g., Wi-Fi Direct). 

MP partitions an LLM into subsets of layers and allocates contiguous subsets of layers across devices (Fig.~\ref{fig:Motivation1}(a)). In the beginning, a device (typically main device held in user hand, such as smartphone) executes the first subset of layers allocated to it. After the device finishes executing the layers, it transmits the output activations to another device that holds the next layers. This process continues sequentially until all layers are executed. MP typically exhibits long computation time and low communication time (Fig.~\ref{fig:Motivation1}(c)), as it requires sequential execution of layers while input/output activations are transmitted only once per device.

TP, on the other hand, splits each layer of an LLM into fine-grained shards and distributes the shards across devices (Fig.~\ref{fig:Motivation1}(b)). In each layer, the main device broadcasts the input activations to the rest of the devices. Then, all the devices simultaneously perform computations of the assigned shards with the input activations. After the devices complete the computation, the results are aggregated to the main device. This process is repeated across all layers to generate output tokens. TP typically exhibits short computation time and high communication time (Fig.~\ref{fig:Motivation1}(c)), due to the concurrent shard execution of layers with frequent communication (i.e., broadcasting and aggregation repeated at every layer).

\begin{rqbox}
\textbf{Takeaway:} MP and TP exhibit distinct computation-communication characteristics, making it difficult to employ one as a one-size-fits-all method.
\end{rqbox}

\begin{figure}[t!]
\centering
\includegraphics[width=0.98\linewidth]{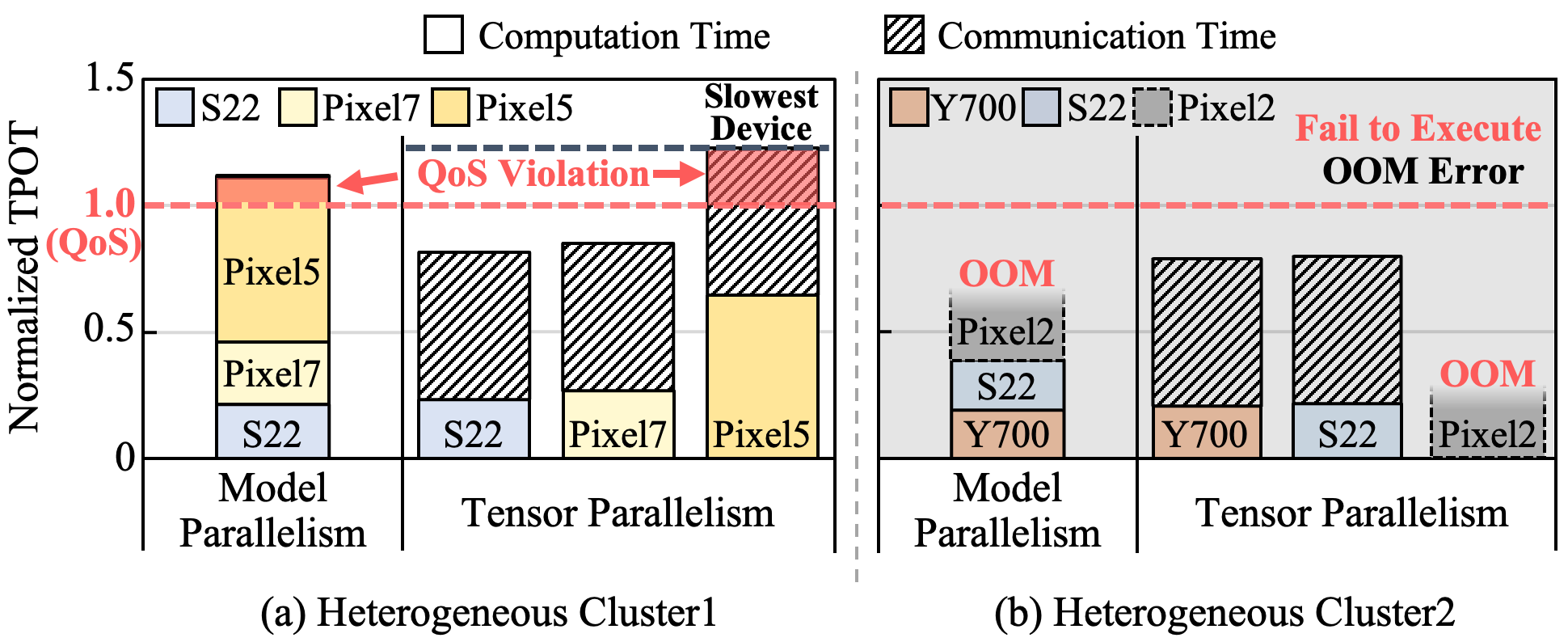}
\caption{TPOT breakdown of multi-device LLM inference under different clusters and parallelism methods.}
\label{fig:Motivation2}
\end{figure}

\vspace{-0.3cm}
\subsubsection{Heterogeneity}
\label{Section3.2.2}

Edge LLM execution inherently exhibits multiple forms of heterogeneity. Edge devices have highly fragmented computational and memory capabilities, introducing a severe degree of performance heterogeneity. A variety of layer architectures and numerical precisions used by LLM also affect the latency and memory footprint of multi device execution, leading to model and precision heterogeneity. In this subsection, we characterize the impact of the heterogeneity on multi-device LLM inference.


\noindent
~\textbf{Device Heterogeneity: } Fig.~\ref{fig:Motivation2} shows TPOT breakdown of multi-device LLM inference under two heterogeneous clusters commonly observed in realistic deployment scenarios\footnote{The detailed configurations of the clusters are explained in Section~\ref{Section5}.}. As shown in Fig.~\ref{fig:Motivation2}, naively adopting MP and TP on heterogeneous clusters fails to provide sufficient execution results. In case of Cluster 1 (Fig.~\ref{fig:Motivation2}(a)), TPOT does not satisfy the QoS requirements. This is because, despite distinct performance capabilities of the edge devices, MP and TP, which are originally designed for identical GPU nodes in large-scale server execution, allocate computations to them evenly --- the device with the lowest computation capability (i.e., Pixel 5 in Fig.~\ref{fig:Motivation2}(a)) thus becomes the performance bottleneck in both MP and TP. Even worse, in case of Cluster 2 (Fig.~\ref{fig:Motivation2}(b)), both MP and TP fails to execute the LLM due to the OOM error. This is also because MP and TP evenly distribute weight tensors and KV cache across the devices without considering the heterogeneous memory budget of the edge devices---the device with the lowest memory budget (i.e., Pixel 2 XL in Fig.~\ref{fig:Motivation2}(b)) cannot accommodate the allocated tensors.

\begin{figure}[t!]
\centering
\includegraphics[width=\linewidth]{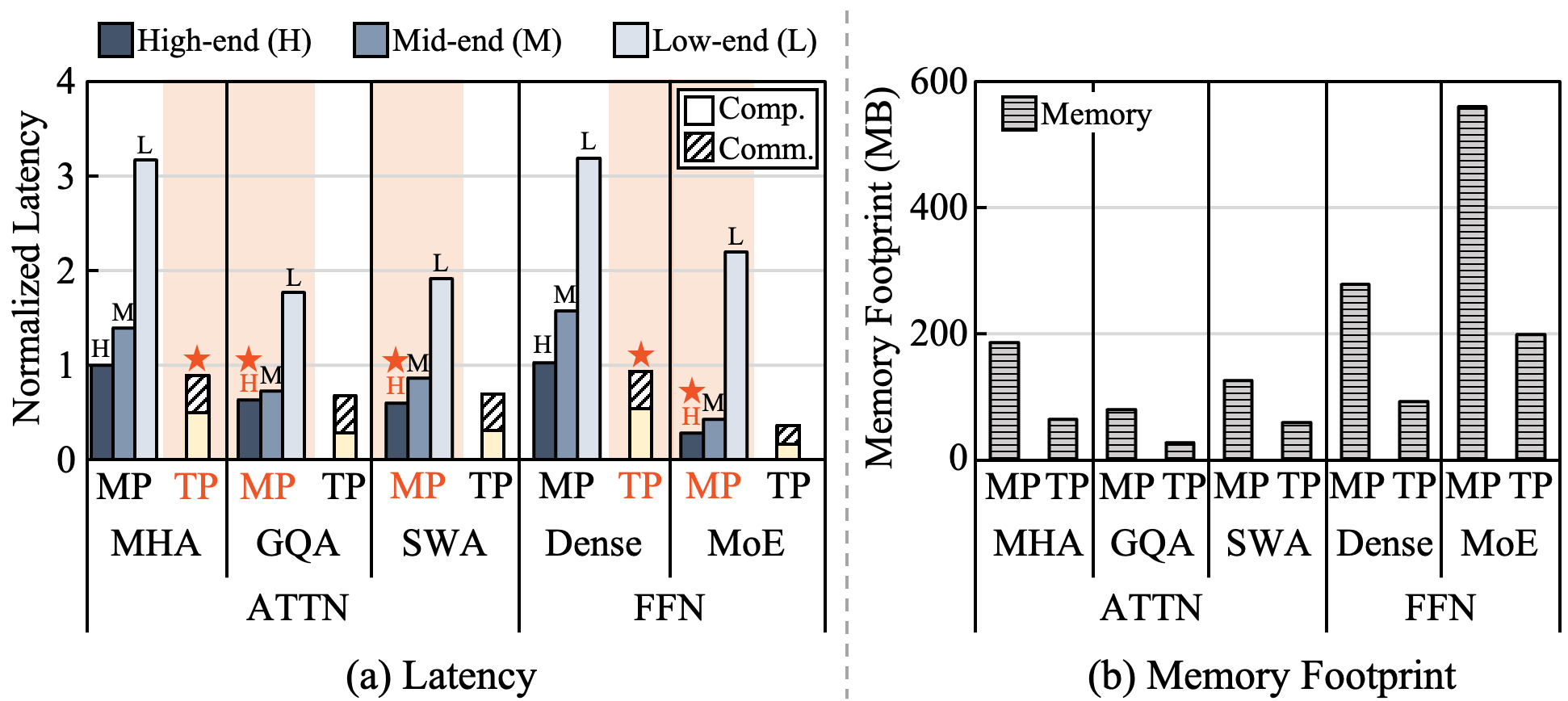}
\caption{Latency and memory footprint of various LLM layers. Note latency is normalized to that of MP in MHA of ATTN.}
\label{fig:Motivation3}
\end{figure}

\begin{rqbox}
\textbf{Takeaway:} Naively employing MP and TP for heterogeneous edge clusters may lead to the failure (and/or QoS violation) of the LLM execution. This calls for the need of new parallelism method tailor-designed for the edge execution environment.
\end{rqbox}


\noindent
~\textbf{Model Heterogeneity: }Recently, state-of-the-art LLMs are adopting hybrid architectures composed of several different ATTN and FFN variants. Such layer variants exhibit distinct performance characteristics when executed with MP and TP on heterogeneous edge devices. Fig.~\ref{fig:Motivation3}(a) shows the latency of various layers when they are allocated to different devices based on MP and TP. As shown in Fig.~\ref{fig:Motivation3}(a), the best performing parallelism strategy varies across the layer types due to their distinct performance characteristics. In case of MHA which has the largest computation load among the ATTN variants, TP shows better performance compared to MP. On the other hand, in case of GQA and SWA which are designed to reduce the computation load of ATTN, MP (on the high-end) shows better performance than TP. Similar patterns can be observed in the FFN variants --- TP outweighs MP for dense FFN while MP outweighs TP for the MoE variant which is designed to exploit sparsity for efficiency.

Those layer variants also exhibit significantly heterogeneous memory footprint on each device, depending on the parallelism strategy (Fig.~\ref{fig:Motivation3}(b)). This makes it difficult to employ per-layer parallelism strategy solely relying on the performance characteristics. For example, in case of MoE variant, MP exhibits higher performance than TP but incurs a significantly larger memory footprint --- among the ATTN variants, GQA and SWA also exhibit similar performance-memory trade-off. This implies that, when choosing parallelism strategies for each layer, it is crucial to consider both performance characteristics and memory requirements.

\begin{rqbox}
\textbf{Takeaway:} Model heterogeneity further complicates the parallelism method selection problem, since different layer variants exhibit distinct performance and memory characteristics under MP and TP.
\end{rqbox}

\begin{figure}[t!]
\centering
\includegraphics[width=\linewidth]{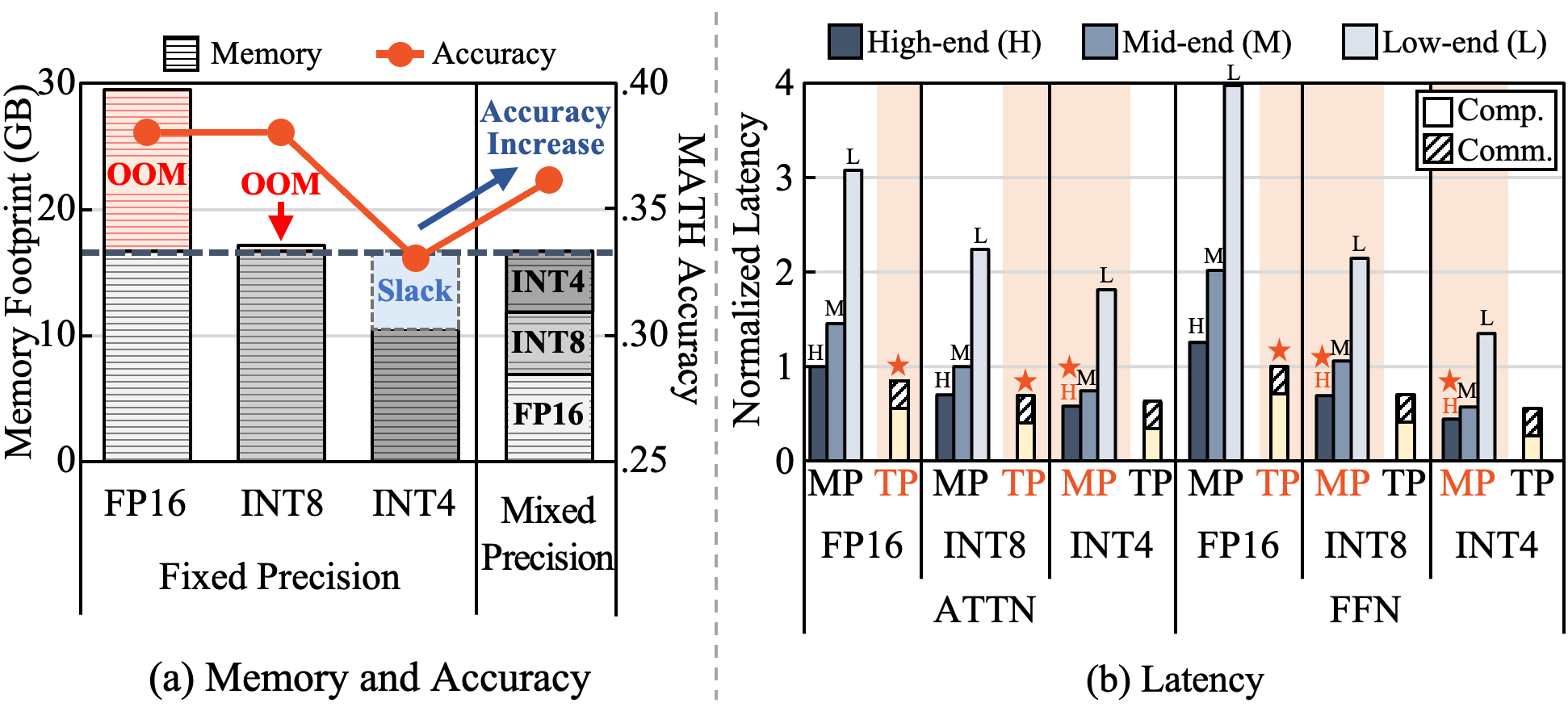}
\caption{(a) Memory usage and accuracy of fixed- and mixed-precision LLMs and (b) layer latency across different precision levels. Note latency is normalized to MP in FP16 of ATTN.}
\label{fig:Motivation4}
\end{figure}


\noindent
~\textbf{Precision Heterogeneity: }To achieve the highest accuracy given limited memory budget, mixed precision is widely used in LLM inference execution~\cite{huang2025slimllmsaliencedrivenmixedprecisionquantization}. Fig.~\ref{fig:Motivation4}(a) compares the accuracy and memory requirements of fixed precision execution and the mixed precision execution with three different precisions (FP16, INT8, and INT4). Higher precisions (i.e., FP16 and INT8) typically exhibit higher accuracy, at the expense of larger memory footprint. Such larger memory footprint often incurs OOM errors on the resource-constrained edge devices. Employing lower precision (i.e., INT4) can mitigate the memory issues, but the accuracy can be significantly degraded. By employing different precisions for each layer (considering its weight distribution and importance), it is possible to achieve reasonably high accuracy, within the limited memory budget. This makes the \textit{precision heterogeneity} natural in the edge execution environment.



Such precision heterogeneity even complicates the parallelism method selection problem. As shown in Fig.~\ref{fig:Motivation4}(b), MP/TP performance characteristics of the same layer can even vary depending on the precision. Lower precision reduces the computational workloads of each shard, increasing the relative impact of communication time in TP execution. Accordingly, the performance benefits of TP becomes smaller at lower precision, while MP remains less affected due to its lower communication overhead. Given that the memory requirements of a layer also scales with the precision, the selection of a parallelism strategy for each layer also requires the consideration of the precision heterogeneity.

\begin{rqbox}
\textbf{Takeaway:} Mixed precision, which is natural in resource-constrained edge environment, introduces additional challenges in the parallelism selection.
\end{rqbox}

\begin{figure}[t!]
\centering
\includegraphics[width=\linewidth]{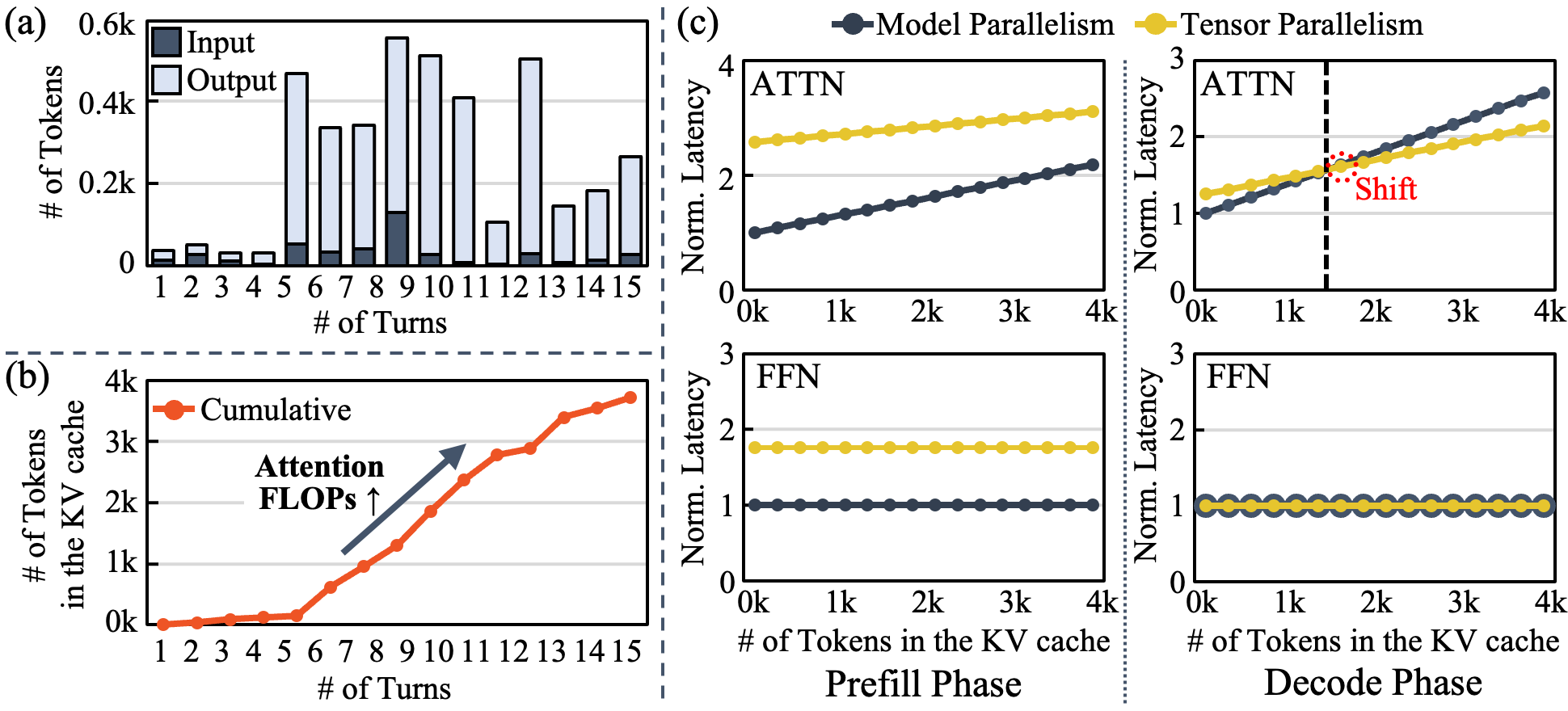}
\caption{(a) Input/output length and (b) the number of tokens accumulated in the KV cache across conversation turns, and (c) layer latency across inference phases and conversation turns. Note latency is normalized to MP with 0k tokens in the KV cache.}
\label{fig:Motivation5}
\end{figure}

\subsubsection{Runtime Variance}
~\label{Section3.2.3}
Edge execution is stochastic by nature. LLM inference itself has computational dynamics due to increasing number of tokens iteratively accumulated into the KV cache, and inference phase changes. Wireless network also exhibits significantly varying signal strength as the user moves. In this subsection, we examine the impact of runtime variance in edge inference execution.

\noindent
~\textbf{Computational Dynamics: }LLM inference exhibits inherent variability in the size of the KV cache, due to the dynamic nature of user conversations. As shown in Fig.~\ref{fig:Motivation5}(a), input and output lengths vary significantly across requests (e.g., standard deviation of input/output length of per-user requests is 160.1/381.7, on average in LMSYS-Chat-1M dataset~\cite{chatbotArena}). This affects the number of tokens accumulated in the KV cache for each token generation. The KV cache size is also gradually enlarged as conversation progresses, as shown in Fig.~\ref{fig:Motivation5}(b). This accumulation increases the computational load of the ATTN, leading to higher latency, as shown in Fig.~\ref{fig:Motivation5}(c). Since TP distributes the growing computational load across devices, its latency increases more slowly than that of MP. As a result, better performing parallelism strategy shifts for ATTN in the decode phase.


The performance characteristics of MP/TP also varies depending on the inference phase. As shown in Fig.~\ref{fig:Motivation5}(c), in the prefill phase, MP achieves lower latency than TP regardless of the number of tokens in KV cache. This is because the prefill phase typically processes many tokens together with batched execution, which diminishes the advantages of TP over MP. On the other hand, in the decode phase, TP can achieve lower latency than MP beyond a certain number of accumulated tokens (right top plot of Fig.~\ref{fig:Motivation5}(c)) --- since the decode phase processes the increased computation load of the ATTN layer with limited batching, TP can better amortize its communication overhead over the computation load, achieving better performance than MP.

\begin{rqbox}
\textbf{Takeaway:} LLM inference has computational dynamics. Since the layer-wise performance characteristics of the parallelism methods can vary with such dynamics, a static parallelism strategy is not sufficient to consistently achieve efficient LLM execution.    
\end{rqbox}

\begin{figure}[t!]
\centering
\includegraphics[width=\linewidth]{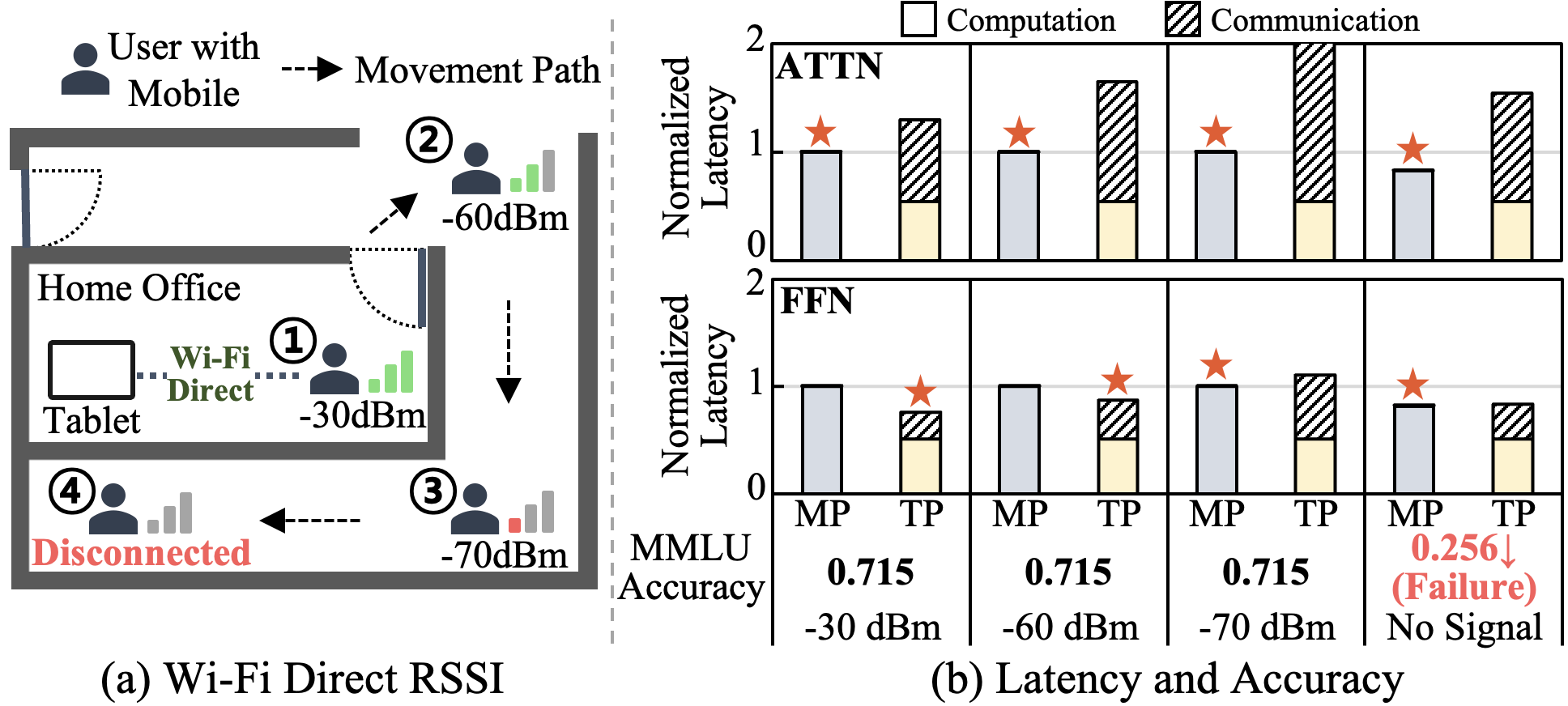}
\caption{(a) Wi-Fi Direct RSSI under user mobility and (b) the performance and accuracy of multi-device LLM execution under varying wireless signal strength. Note latency is normalized to MP with -30 dBm of signal strength.}
\label{fig:Motivation6}
\end{figure}

\noindent
\textbf{Stochastic Wireless Network: }Network variability also affects communication overheads, shifting the performance characteristics of the per-layer parallelization strategies. Fig.~\ref{fig:Motivation6}(a) shows an example of the sources of network variability. As a user moves within the smart home environment, the distance between the devices changes, affecting the wireless signal strength of their wireless connections. Such signal strength variation significantly affects the communication overhead, as shown in Fig.~\ref{fig:Motivation6}(b). In case of ATTN, MP consistently outweighs TP due to the relatively small computation load --- the communication overhead does not account for a large portion of total execution time. In contrast, in case of FFN, the performance characteristics of MP/TP vary with the wireless signal strength. When signal strength is sufficiently strong (e.g., up to -60 dBm), TP achieves better latency than MP. However, when the Wi-Fi Direct RSSI drops to -70dBm, TP incurs substantial communication overhead due to bandwidth limitations and increased latency. In such cases, MP becomes a better option. When the Wi-Fi Direct signal strength falls below a certain threshold, the connection can be even lost, as illustrated in Fig.~\ref{fig:Motivation6}(a). In this case, the loss of connectivity prunes the weight tensors assigned to the disconnected device, degrading accuracy (Fig.~\ref{fig:Motivation6}(b)).

\begin{figure*}[t!]
\includegraphics[width=0.98\linewidth]{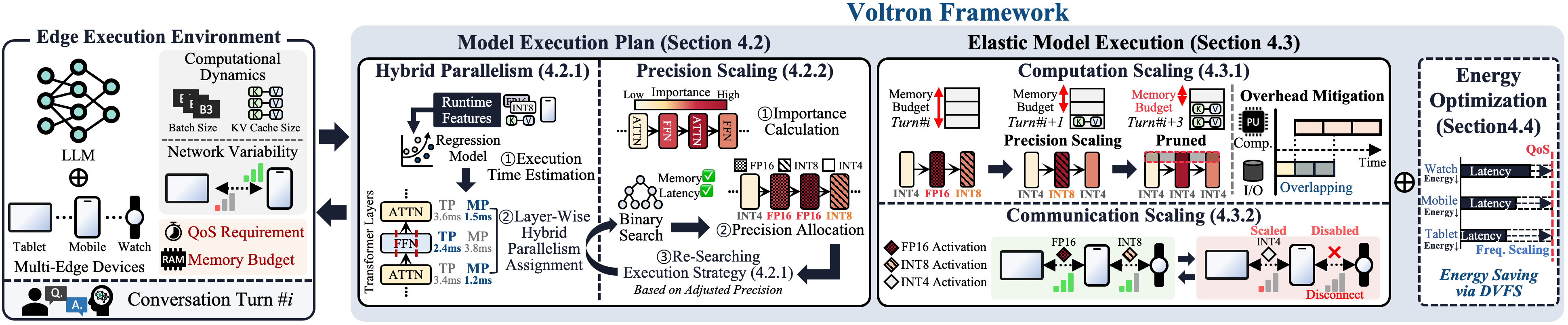}
\vspace{-0.3cm}
\caption{Overview of Voltron Framework.}
\label{fig:Design1}
\end{figure*}

\begin{rqbox}
\textbf{Takeaway:} The wireless network variability also needs to be carefully considered, when deciding the parallelism strategies.
\end{rqbox}

\vspace{-0.2cm}
\section{Voltron}~\label{Section4}
\vspace{-0.2cm}

\vspace{-0.3cm}
\subsection{Overview}
~\label{Section4.1}
To address the challenges of edge execution, we propose a novel multi-device LLM inference framework, ~\textbf{Voltron} (Fig.~\ref{fig:Design1}), which enables elastic execution of LLM across edge devices. At the beginning of each conversation turn, Voltron determines model execution plan, including layer-wise parallelism strategy and precision (Section~\ref{Section4.2}). For overcoming the limitations of the conventional parallelism methods, Voltron employs a novel \textit{hybrid parallelism (HP)} method, tailor-designed for edge execution environment. During the LLM execution, Voltron continuously monitors execution environment, and elastically adjusts the execution strategy by adapting to the runtime variance (Section~\ref{Section4.3}).

\vspace{-0.2cm}
\subsection{Model Execution Plan}
~\label{Section4.2}
At the beginning of each conversation turn, Voltron determines a model execution plan that specifies the layer-wise parallelism strategy and precision configuration, to maximize accuracy while satisfying QoS requirements under the current execution environment. Voltron first searches for the fastest parallelism strategy for each layer based on the estimated execution time (Section~\ref{Section4.2.1}) while meeting the memory budget. It then adjusts the precision configurations of each layer to better satisfy the QoS requirements while maximizing accuracy (Section~\ref{Section4.2.2}). Since layer-wise performance characteristics of MP/TP can differ across inference phases, as discussed in Section~\ref{Section3.2.3}, it explores separate execution strategies for the prefill and decode phases.

\subsubsection{Layer-wise Hybrid Parallelism}
~\label{Section4.2.1}
Voltron sequentially assigns a parallelism strategy (i.e., MP or TP) to each layer, while considering the memory budget of each device. For each layer, Voltron estimates the execution time (i.e., sum of computation and communication time) of both strategies and selects the one with lower latency. The computation time is estimated exploiting a regression model (of which average accuracy is 90\% that is sufficient to correctly identify the better-performing parallelism method) based on runtime features such as device type, layer type, precision, input token length, and the number of tokens accumulated in the KV cache~\cite{10.1145/3695053.3730999}. The communication time is estimated based on the size of transmission data and wireless network conditions (e.g., available network bandwidth and signal strength)~\cite{neurosurgeon}.


Voltron determines the layer allocation for the selected parallelism strategy as follows. (1) MP: Voltron first attempts to place the layer on the highest-performance device to minimize execution time. If the layer does not fit within the device's memory budget, Voltron selects the next available device that satisfies the memory requirement, while having the best performance. (2) TP: Voltron greedily assigns shards (i.e., partitions of a layer) to devices based on their performance to balance the execution time across devices, thereby minimizing the impact of stragglers.

\subsubsection{Importance-aware Mixed Precision}
~\label{Section4.2.2}
After the execution strategy is determined, Voltron adjusts the precision for each layer to maximize accuracy while satisfying latency and memory requirements. Since each layer has different impact on inference accuracy~\cite{he2024matterstransformersattentionneeded}, Voltron first profiles the accuracy impact of each layer in an offline phase. Specifically, Voltron calculates an importance score for each layer using channel-level importance metric~\cite{michel2019sixteenheadsreallybetter}. For each layer, we first evaluate the perplexity of the full model with the highest precision (i.e., FP16), by using the validation dataset. After that, we re-evaluate the perplexity while pruning channels within the layer one by one. The perplexity difference measured while pruning each channel is then identified as the importance score of the channel---the higher difference is, the larger its importance is. We then add up the importance scores of all channels within each layer to identify the importance score of the layer. Since layer importance patterns can vary across tasks, we perform the offline profiling separately for each task. Given that the precision changes can affect the layer-wise performance characteristics, Voltron re-searches for the better execution strategy with the execution plan module (Section~\ref{Section4.2.1}) based on the adjusted precision configuration.


When allocating precisions to the layers by considering the layer-wise parallelism strategies and the memory budget, it is infeasible to enumerate all the possible allocations at runtime especially in the resource-constrained edge devices --- the complexity of the enumeration is exponential in the number of layers in the worst case. To efficiently explore the precision allocations by exploiting the monotonic relationship between the precision, memory usage, and inference latency (that we observe in Section~\ref{Section3.2.2}), Voltron uses binary search of which complexity is $O\log(n)$ --- quantitative overhead analysis is presented in Section~\ref{Section5}. Voltron performs the binary search starting from the full-precision (i.e., FP16) model and progressively scales layer precision in the ascending order of importance. At each step, it constructs a candidate mixed-precision model and checks whether the resulting configuration satisfies both the memory budget and QoS requirements. Then, it adjusts the search direction accordingly until the final configuration is determined.


\subsection{Elastic Model Execution}
~\label{Section4.3}
As we observe in Section~\ref{Section3.2.3}, edge execution of LLM inference is inherently stochastic due to computational dynamics and wireless network variability. To adapt to the runtime variance, Voltron continuously adjusts the execution strategy. Specifically, Voltron adapts to the computational dynamics of LLM inference by adjusting the execution plan (Section~\ref{Section4.3.1}). Voltron also adapts to the wireless network variability via selective activation quantization (Section~\ref{Section4.3.2}).

\subsubsection{Computation Scaling}
~\label{Section4.3.1}
To adapt to computational dynamics, Voltron first monitors the KV cache size and the currently available memory budget\footnote{In our implementation, the communication scaling module is executed in advance to the computation scaling module in order to prevent communication overhead being a major performance bottleneck --- the QoS margin is already minimized by the communication overhead in such case, so that computation scaling has a limited accuracy protection room. Note, if some shards in layers are already pruned by communication scaling, the computation scaling module preserves the pruning results and checks the memory budget based on the pruned architecture.} for each token generation. When the memory budget is expected to be insufficient for the next token generation, due to the increase of KV cache size or device disconnection, Voltron first adopts precision scaling. If solely adopting precision scaling is not sufficient to resolve the memory budget shortage, Voltron additionally adopts selective pruning --- as we demonstrate in Fig.~\ref{fig:Background2}(a) pruning results in more severe accuracy drop compared to the adopting INT4 precision to entire model. When memory budget headroom becomes available again (e.g., due to KV cache flush or connection recovery), Voltron re-executes the model execution plan module (Section~\ref{Section4.2}) to gradually recover the layers and their precisions.

\noindent\textbf{Precision Scaling:}
~\label{Section4.3.1.1}
The goal of precision scaling is to find the best layer-wise precision that meets the memory budget and QoS constraint while minimizing the accuracy loss. To achieve the goal, Voltron re-executes the model execution plan module to generate new execution plan (i.e., precision-aware layer-wise parallelism strategy) the most suitable for the changed execution environment --- the overhead of the model execution plan module only accounts for 0.2\% of TPOT (i.e., less than a few hundred microseconds even in the worst case), thereby making it feasible to run the module for token-by-token modifications. Once the new execution plan is determined, Voltron updates only the layers that require precision scaling. Since these updates may incur non-trivial I/O overhead, we discuss an approach to mitigate this overhead in \textbf{Overhead Mitigation}.

\noindent\textbf{Pruning:}
~\label{Section4.3.1.2}
Since pruning may incur severe accuracy drop, Voltron adopts importance-aware pruning strategy, which follows the similar principle used in an importance-aware precision allocation (Section~\ref{Section4.2.2}). Specifically, it adopts shard-wise pruning in order to minimize the information loss. Voltron first determines the number of pruning shards based on the total memory budget. Note Voltron prunes the same number of shards from all layers, to ensure the balanced loss across the layers --- this practice has been widely adopted in a number of pruning works to minimize the accuracy loss~\cite{LLMPruner}. Voltron then prunes the shards in the ascending order of importance score for each layer. After that, Voltron modifies the execution plan based on the newly estimated execution time of per-layer parallelization strategies and memory budget of respective devices, by incorporating the pruned layers. Each device reflects the plan modification by 1) reloading the newly allocated layers from the storage and 2) removing the pruned shard weights from the layers allocated in the memory if needed. The overhead of the former can be mitigated by the \textbf{Overhead Mitigation} approach while that of the latter can be mitigated via shard-wise memory allocations --- the pruned shard weights can easily be removed from the layers allocated in the memory, and/or from the layers that are newly allocated for the replacement.

\noindent\textbf{Overhead Mitigation:}
When Voltron replaces shards (or layers) stored in the memory with new ones from storage, additional I/O operations are required. Performing I/O operations during the LLM inference can severely degrade performance due to the limited bandwidth of mobile storage (e.g., UFS). To minimize the I/O overhead, Voltron overlaps I/O operations with ongoing LLM computation by exploiting idle computing resources within each device. Specifically, while executing the LLM computations, Voltron utilizes idle computing resources to preload the new shards (or layers) from storage. By overlapping these data transfers with LLM computations, Voltron effectively hides most of the I/O latency --- the detailed analysis on the effectiveness of the overhead mitigation is presented in Section~\ref{Section5.2.4}.

\subsubsection{Communication Scaling}
~\label{Section4.3.2}
To adapt to the wireless network variability, Voltron estimates the expected communication overhead for each token generation under the current wireless network signal strength. When the expected communication overhead becomes too large to push the LLM latency beyond the QoS requirement, Voltron determines how much the cross-device transmission data needs to be reduced to satisfy the QoS constraint. To achieve this reduction, Voltron lowers the precision of activations (required to be transmitted across devices) via quantization. Since activation quantization may incur accuracy loss, Voltron employs an importance-aware approach. Specifically, Voltron computes the importance of each activation as the sum of the importance of the weight shards that are used for generating it, and lowers the precision of activations in the ascending order of importance score. If lowering all activations to the lowest precision (i.e., INT4) is still insufficient to achieve the required reduction in the amount of transmitted data, Voltron further decreases communication by progressively pruning transmission activations (as well as the corresponding shards) based on their importance. When the wireless network status improves, Voltron gradually restores the pruned activations and their precisions, to recover the model accuracy.

\begin{table}[t]
\caption{Mobile Device Specification}
\vspace{-0.2cm}
\label{table1}
\resizebox{\columnwidth}{!}{
\begin{tabular}{|c|c|c|c|}
\hline
\textbf{Device Name}& \textbf{DRAM / Available} & \textbf{SoC} & \textbf{Performance Level} \\ \hline
Lenovo Legion Y700  & 12GB / 8.3GB & Snapdragon 8+ Gen 1       & High-end \\ \hline
Galaxy S22 Ultra    & 12GB / 7.3GB & Snapdragon 8 Gen 1        & High-end \\ \hline
Google Pixel 7      & 8GB / 4.7GB  & Google Tensor G2          & Mid-end  \\ \hline
Google Pixel 5      & 8GB / 4.7GB  & Snapdragon 765G SM7250    & Mid-end  \\ \hline
Google Pixel 2 XL   & 4GB / 2.3GB  & Snapdragon 835 MSM8998    & Low-end  \\ \hline
Nexus 5X            & 2GB / 1.2GB  & Snapdragon 808 MSM8992    & Low-end  \\ \hline
\end{tabular}%
}
\end{table}

\subsection{Energy Optimization}
~\label{Section4.3.3}
Although the multi-device LLM inference improves model accuracy while satisfying QoS requirements, it may increase the energy consumption of the participating devices. If the devices are all battery-powered, alleviating the energy increase might get crucial. To minimize the energy consumption in such cases, Voltron includes an energy optimization module. When energy saving is prioritized, the energy module further reduces latency of LLM inference by employing computation scaling (i.e., precision scaling and/or pruning). This creates a latency margin (i.e., the gap between the reduced latency and the latency constraint). The energy optimization module then exploits this margin to reduce the power consumption. Specifically, it reduces the voltage and frequency of the processing units of each device, as long as the latency constraint is satisfied --- the effectiveness of the energy module is presented in Section~\ref{Section5.2.3}.

\begin{table}[t]
\centering
\caption{Cluster of Devices}
\label{table2}
\resizebox{\columnwidth}{!}{%
\begin{tabular}{|c|c|c|c|}
\hline
\textbf{Cluster} & \textbf{Device1} & \textbf{Device2} & \textbf{Device3} \\ \hline

\begin{tabular}[c]{@{}c@{}}Car\\(Cluster1)\end{tabular}
& \begin{tabular}[c]{@{}c@{}}Mobile\\(Galaxy S22 Ultra)\end{tabular}
& \begin{tabular}[c]{@{}c@{}}Automotive SoC\\(Google Pixel 7)\end{tabular}
& \begin{tabular}[c]{@{}c@{}}Infotainment SoC\\(Google Pixel 5)\end{tabular}
\\ \hline

\begin{tabular}[c]{@{}c@{}}Office\\(Cluster2)\end{tabular}
& \begin{tabular}[c]{@{}c@{}}Mobile\\(Galaxy S22 Ultra)\end{tabular}
& \begin{tabular}[c]{@{}c@{}}Tablet\\(Lenovo Legion Y700)\end{tabular}
& \begin{tabular}[c]{@{}c@{}}Smart Watch\\(Google Pixel 2 XL)\end{tabular}
\\ \hline

\begin{tabular}[c]{@{}c@{}}Home\\(Cluster3)\end{tabular}
& \begin{tabular}[c]{@{}c@{}}Mobile\\(Galaxy S22 Ultra)\end{tabular}
& \begin{tabular}[c]{@{}c@{}}Smart Watch\\(Google Pixel 2 XL)\end{tabular}
& \begin{tabular}[c]{@{}c@{}}IoT Device\\(Nexus 5X)\end{tabular}
\\ \hline

\end{tabular}
}
\end{table}

\begin{figure*}[t!]
\includegraphics[width=\linewidth]{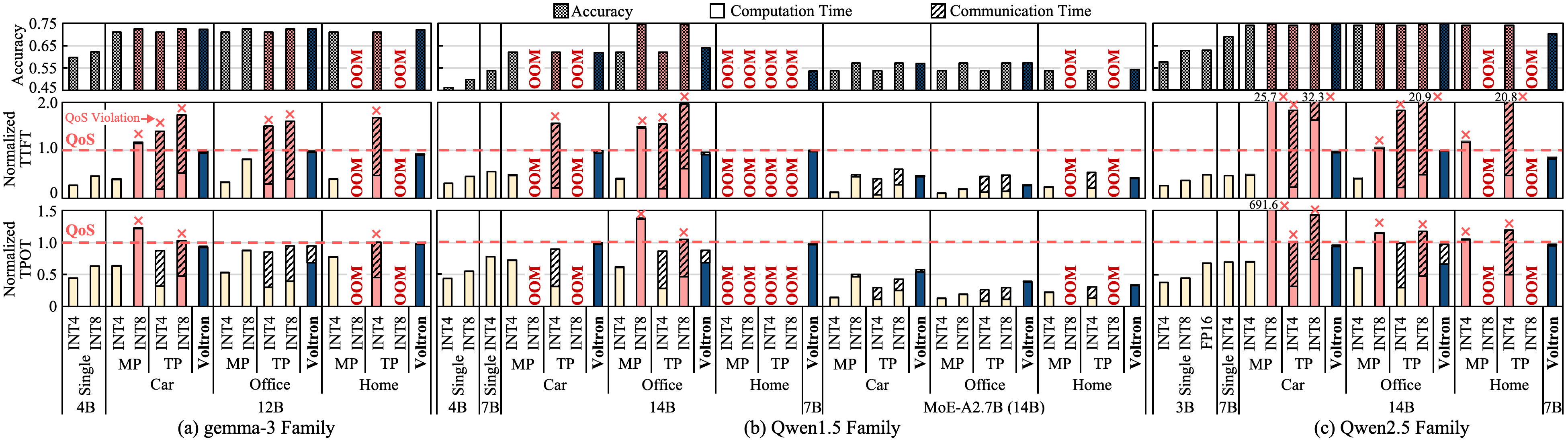}
\caption{Normalized TTFT/TPOT and accuracy for Voltron and baselines across clusters and models.}
\label{fig:Evaluation1}
\end{figure*}

\section{Evaluation}
\label{Section5}

\subsection{Experimental Setup}
\label{Section5.1}

\subsubsection{Devices}
\label{Section5.1.1}
We perform experiments on six mobile devices spanning a wide range of performance levels and memory budgets: two high-end devices representing tablets and smartphones, two mid-end devices representing automotive and infotainment SoCs, and two low-end devices representing smartwatches and IoT devices. Table~\ref{table1} summarizes their specifications. We configure three clusters with the devices (Table~\ref{table2}), reflecting realistic use case scenarios~\cite{samsungsmarthings,tesla,IoTscenario}: Cluster 1 (Car), Cluster 2 (Office), and Cluster 3 (Home). In each cluster, the devices are connected via Wi-Fi Direct --- other wireless network, such as Bluetooth, can also be used.

\subsubsection{Workloads}
\label{Section5.1.2}
We use three open-source state-of-the-art LLM families (i.e., gemma-3~\cite{gemmateam2025gemma3technicalreport}, Qwen1.5~\cite{bai2023qwentechnicalreport}, and Qwen2.5 \cite{qwen2025qwen25technicalreport}) with various parameter sizes and layer types (Table~\ref{table3}). To support the mixed precision, we adopt k-quant~\cite{k-quant}, which provides implementation of multiple precision formats including FP16, INT8, and INT4. We use input traces of LMSYS-Chat-1M dataset~\cite{chatbotArena} to emulate realistic conversational workloads along with varying input/output length. We use representative LLM benchmarks for accuracy evaluation: MMLU (0-shot), Hellaswag (0-shot), GSM8K (8-shot, CoT), and MATH (4-shot, CoT)~\cite{hendrycks2021measuringmassivemultitasklanguage, zellers2019hellaswagmachinereallyfinish, cobbe2021trainingverifierssolvemath, hendrycks2021measuringmathematicalproblemsolving}.

\begin{table}[t]
\centering
\caption{Large Language Models}
\vspace{-0.2cm}
\label{table3}
\resizebox{\columnwidth}{!}{%
\begin{tabular}{|c|c|c|c|c|}
\hline
~\textbf{Family}                   & ~\textbf{Nam}e              & ~\textbf{\#   of Parameters} & ~\textbf{ATTN}           & ~\textbf{FFN}         \\ \hline
\multirow{2}{*}{gemma-3} & gemm-3-4b-it      & 4B                 & GQA, SWA(1024) & Dense FFN   \\ \cline{2-5} 
                         & gemm-3-12b-it     & 12B                & GQA, SWA(1024) & Dense FFN   \\ \hline
\multirow{4}{*}{Qwen1.5} & Qwen1.5-4B        & 4B                 & MHA            & Dense   FFN \\ \cline{2-5} 
                         & Qwen1.5-7B        & 7B                 & MHA            & Dense   FFN \\ \cline{2-5} 
                         & Qwen1.5-14B       & 14B                & MHA            & Dense   FFN \\ \cline{2-5} 
                         & Qwen1.5-MoE-A2.7B & 14B                & MHA            & MoE FFN     \\ \hline
\multirow{3}{*}{Qwen2.5} & Qwen2.5-4B        & 3B                 & GQA            & Dense   FFN \\ \cline{2-5} 
                         & Qwen2.5-7B        & 7B                 & GQA            & Dense   FFN \\ \cline{2-5} 
                         & Qwen2.5-14B       & 14B                & GQA            & Dense   FFN \\ \hline
\end{tabular}
}
\end{table}

\subsubsection{Execution Scenarios} 
\label{Section5.1.3}
We implement Voltron on top of open-source LLM framework \textit{llama.cpp}, incorporating the proposed model execution plan and elastic model execution. To evaluate the effectiveness of Voltron, we compare it against 1) single-device LLM execution\footnote{For the single-device execution, we run the LLM on the highest-performing mobile device in the cluster.} and 2) device heterogeneity-aware MP/TP~\cite{PerformanceAwareMP, performanceAwareTP}, which allocates shards or layers based on device performance while considering memory budget, under various execution scenarios. We evaluate the performance using two latency metrics: TTFT and TPOT. Following prior works~\cite{sageserve, 10.1007/978-3-030-23535-2_30}, we adopt QoS targets of 10s and 400ms for TTFT and TPOT, respectively. We evaluate the accuracy on the benchmark tasks described in Section~\ref{Section5.1.2}. We also evaluate energy consumption of the devices by using an external power measurement device~\cite{Monsoon}.

\subsubsection{Runtime Variance Scenarios} 
To demonstrate that Voltron can maintain high accuracy while satisfying the QoS requirements under runtime variance, we construct experimental scenarios that emulate realistic edge execution conditions and evaluate the performance and accuracy during multi-device LLM inference. For computational dynamics, we conduct experiments on Cluster 2 (Office) while increasing the number of conversation turns. For network variability, we emulate a scenario (depicted in Fig.~\ref{fig:Motivation6}(a) in Section~\ref{Section3.2.3}) where a user moves within an office while carrying a mobile device and a smartwatch, with a tablet placed on a desk. In this setup, we collect cross-device wireless signal strength variations from the real deployment environment.

\begin{table}[t]
\centering
\vspace{-10pt}
\caption{Accuracy on LLM Benchmarks}
\label{table4}
\resizebox{\columnwidth}{!}{%
\renewcommand{\arraystretch}{1.1}
\setlength\tabcolsep{2.5pt}
\fontsize{7.5}{7.5}\selectfont

\begin{tabular}{c|c|cc|c|cc|cc|c|cc}
\toprule
& \multicolumn{3}{c|}{\textbf{gemma-3 Family}}
& \multicolumn{5}{c|}{\textbf{Qwen1.5 Family}}
& \multicolumn{3}{c}{\textbf{Qwen2.5 Family}} \\
\midrule

& \multicolumn{1}{c|}{4B}
& \multicolumn{2}{c|}{12B}
& \multicolumn{1}{c|}{7B}
& \multicolumn{2}{c|}{14B}
& \multicolumn{2}{c|}{MoE-A2.7B (14B)}
& \multicolumn{1}{c|}{7B}
& \multicolumn{2}{c|}{14B} \\

\cmidrule(lr){2-2}
\cmidrule(lr){3-4}
\cmidrule(lr){5-5}
\cmidrule(lr){6-7}
\cmidrule(lr){8-9}
\cmidrule(lr){10-10}
\cmidrule(lr){11-12}

\textbf{Benchmark}

& Single 
& MP/TP & \textbf{Voltron}

& Single
& MP/TP & \textbf{Voltron}
& MP/TP & \textbf{Voltron}

& Single
& MP/TP & \textbf{Voltron}

\\
\midrule

\textbf{MMLU}
& 0.5753
& 0.7158 & 0.715 
& 0.5748
& 0.654 & 0.6586
& 0.6094 & 0.6087
& 0.7047
& 0.7693 & 0.7747 \\

\textbf{Hellaswag}
& 0.7428
& 0.8195 & 0.8191
& 0.7658
& 0.7905 & 0.7936
& 0.773  & 0.7727 
& 0.7818
& 0.8223 & 0.8285 \\

\textbf{GSM8K}
& 0.7703
& 0.8946 & 0.8946 
& 0.5959
& 0.7051 & 0.7324
& 0.6262 & 0.6164
& 0.8211
& 0.862 & 0.8681 \\

\textbf{MATH}
& 0.4044
& 0.4752 & 0.476
& 0.215
& 0.332 & 0.377
& 0.2784 & 0.2754 
& 0.4536
& 0.5162 & 0.5226 \\

\midrule
\textbf{Average}
& 0.6232
& 0.7263 & 0.7262
& 0.5379
& 0.6204 & 0.6404
& 0.5718 & 0.5683
& 0.6903
& 0.7425 & 0.7485 \\

\bottomrule
\end{tabular}
}
\vspace{-10pt}
\end{table}

\subsection{Experimental Results and Analysis}
\label{Section5.2}

\subsubsection{Result Overview}
\label{Section5.2.1}

Fig.~\ref{fig:Evaluation1} shows the TTFT/TPOT normalized to the respective QoS constraints, and average accuracy on various clusters of edge devices with three LLM families. Across all clusters and models, Voltron satisfies the QoS requirements for both TTFT and TPOT, as it allocates appropriate parallelism method to each layer with \textit{hybrid parallelism}, by carefully considering the cluster capabilities (i.e., memory budgets and compute capabilities), model architecture, and the computational characteristics that vary across inference phases --- the detailed analysis on the parallelism allocation is presented in Section~\ref{Section5.2.2}. In addition, Voltron achieves significantly higher accuracy compared to the baselines. The accuracy gain comes from two key factors: 1) the ability to execute larger LLMs by leveraging multiple devices and 2) the use of importance-aware mixed precision, which preserves higher precision for importance layers (improving accuracy by up to 1.8\% as in Table~\ref{table4})  as long as the memory budget allows. Voltron achieves 10.7\% and 1.0\% higher average accuracy than single-device execution and MP/TP, respectively. Note, although Voltron allocates higher precision than MP/TP on average, the validation accuracy is slightly lower in some cases --- the difference is less than 0.3\% though. MP/TP, however, are not satisfying the QoS constraints (or even result in execution failure) in those cases while Voltron robustly satisfies the QoS constraints.

\subsubsection{Adaptability Analysis}
\label{Section5.2.2}
\begin{figure}[t!]
\includegraphics[width=\linewidth]{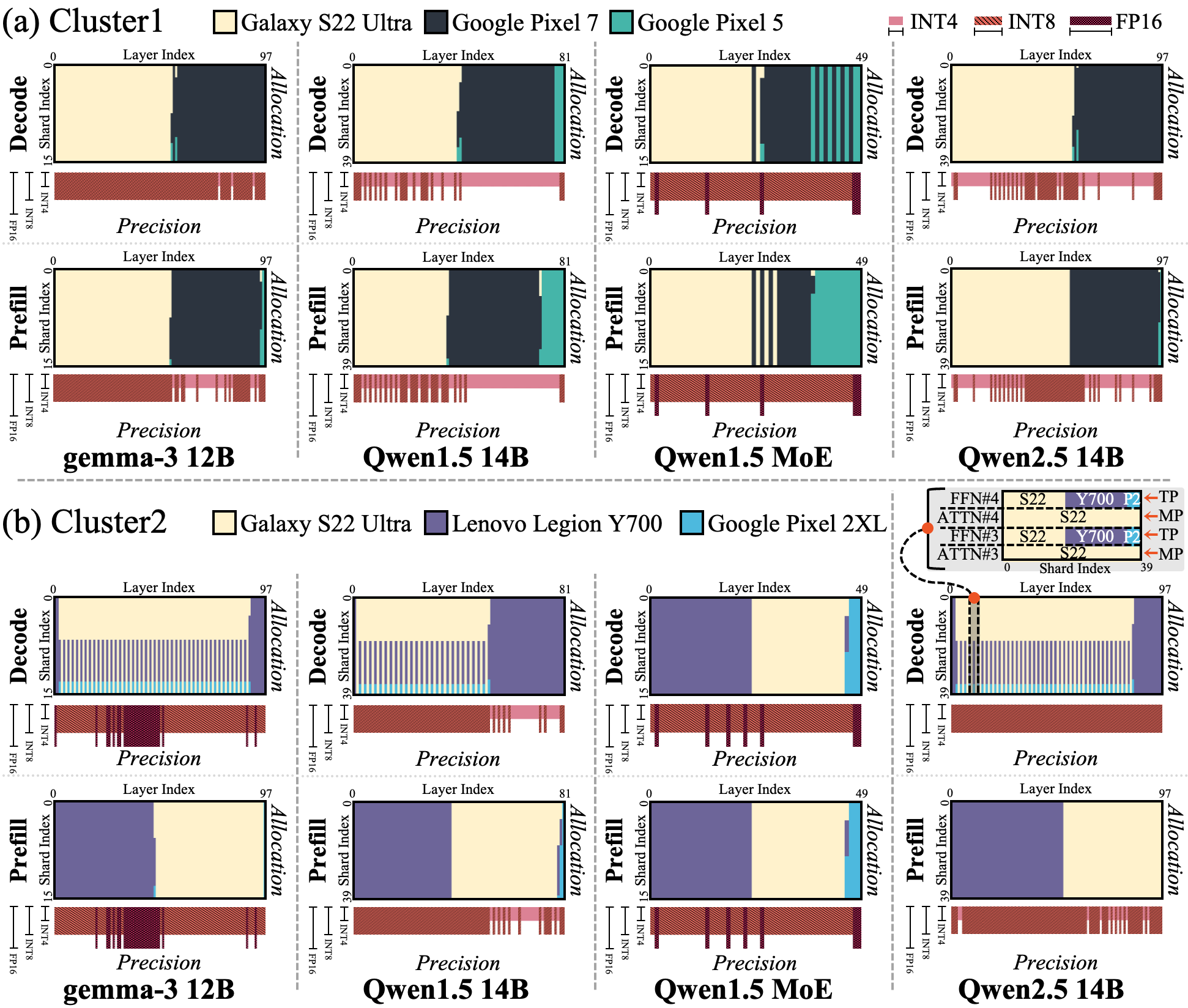}
\caption{Execution strategies determined by Voltron on the Cluster 1 and Cluster 2 across different models. Note Qwen2.5 14B Decode Phase in Cluster2 informs which layers are allocated with MP/TP.}
\label{fig:Evaluation2}
\end{figure}

\textbf{Adaptability to Device and Model Heterogeneity: }            
Voltron can adapt to device and model heterogeneity, by selecting efficient hybrid parallelism strategies which satisfy the latency constraints and memory budget. Fig.~\ref{fig:Evaluation2} shows the execution strategies used by Voltron across two heterogeneous clusters and four model types. Voltron determines the execution plans tailored to heterogeneous device capabilities across clusters and model types. For example, in Cluster 1 (Fig.~\ref{fig:Evaluation2}(a)), most layers are executed with MP in the decode phase whereas some layers are executed with TP in Cluster 2 (Fig.~\ref{fig:Evaluation2}(b)). This is because performance gap across devices is higher in Cluster 1, compared to Cluster 2, which makes TP less performing compared to MP due to the severe straggler problem. Voltron also executes each layer with an appropriate parallelism method according to the layer type. As shown in Fig.~\ref{fig:Evaluation2}(b), Voltron executes most ATTNs, which have relatively low computational load, with MP, while executing most FFNs, which have relatively high computational load, with TP --- in case of Qwen1.5-MoE-A2.7B, on the other hand, most FFNs are executed with MP due to sparse computation of the MoE.

\begin{figure}[t!]
\includegraphics[width=\linewidth]{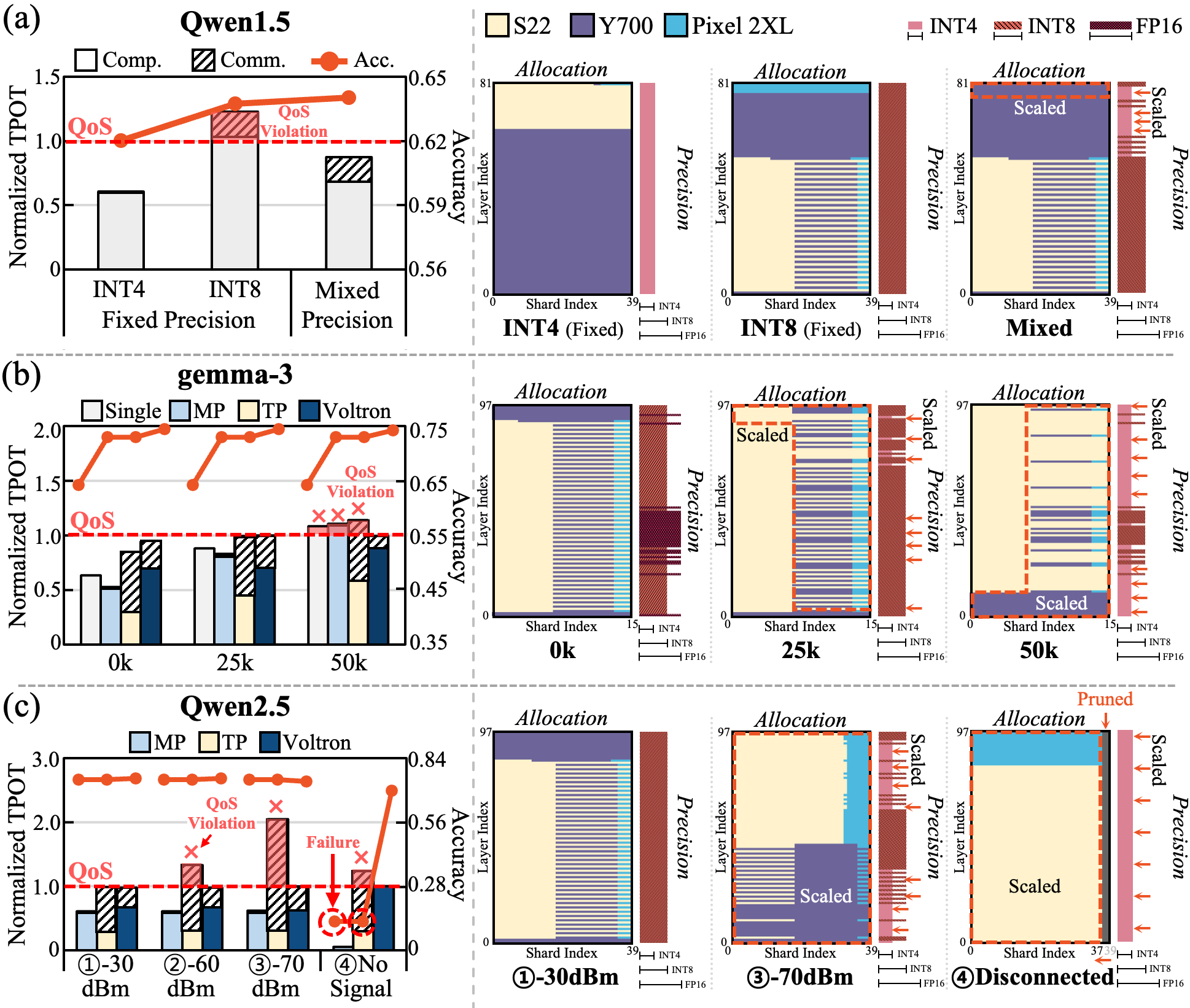}
\caption{Adaptability to (a) precision heterogeneity, (b) computation dynamics, and (c) network variability.}
\label{fig:Evaluation3}
\end{figure}

\noindent
\textbf{Adaptability to Precision Heterogeneity: }  
By using mixed precision, Voltron can maximize accuracy while satisfying latency constraints. Fig.~\ref{fig:Evaluation3}(a) shows the accuracy and latency under fixed- and mixed-precision for Qwen1.5, along with the execution strategy determined by Voltron. In the INT4 fixed-precision setting, Voltron satisfies the latency constraints but exhibits lower accuracy due to information loss. In the INT8 fixed-precision setting, Voltron achieves higher accuracy but fails to satisfy the QoS latency requirements due to the increased computational load. In contrast, by using mixed precision, Voltron achieves higher accuracy while satisfying the QoS latency constraints by effectively exploiting the latency-accuracy trade-off across precisions --- as shown in Fig.~\ref{fig:Evaluation3}(a), Voltron reduces memory usage by lowering the precision of certain layers, while still allocating high precisions to important layers, thereby achieving higher accuracy within the QoS constraint.

\noindent
\textbf{Adaptability to Computational Dynamics: }
Voltron can adapt to computational dynamics. Fig.~\ref{fig:Evaluation2}(b) shows that Voltron employs different execution strategies across inference phases. In the prefill phase, Voltron uses MP for most layers regardless of the cluster and model type. This is because batched execution reduces the per-token computation time, amortizing the overhead of MP’s sequential execution. In contrast, during the decode phase, Voltron appropriately assigns MP or TP to each layer according to the execution environment.


Voltron can also handle large KV cache size (i.e., the number of tokens accumulated in the KV cache) while satisfying latency constraints and maintaining high accuracy. Fig.~\ref{fig:Evaluation3}(b) presents the accuracy and latency of the baselines and Voltron under the varying KV cache size increases. When the KV cache size is 0k, all baselines and Voltron satisfy the latency constraints because the computational load of ATTN is relatively low. However, when the KV cache size increases to 50k, none of the baselines satisfy the latency constraints, whereas Voltron still meets the latency constraints through computation scaling while achieving high accuracy. This is because 1) the importance-aware precision allocation enables Voltron to preserve high accuracy even when most layers are scaled to INT4 due to the reduced memory budget, and 2) the execution plan module appropriately generates layer-wise parallelism strategies for the heterogeneous precisions meeting the QoS constraint. This result implies that Voltron can strike the balanced memory-latency-accuracy trade-off point even for the long context applications. 

\noindent
\textbf{Adaptability to Network Variability: }
Voltron maintains high accuracy satisfying the latency constraints under various wireless signal strength. We consider a scenario where a table is located in a room while a user carrying a watch and a mobile devices moves away from the tablet (which is illustrated in Fig.~\ref{fig:Motivation6}(a) of Section~\ref{Section3.2.3}). Fig.~\ref{fig:Evaluation3}(c) shows the accuracy and latency of baselines and Voltron as the distance between the user and the tablet changes in the scenario. When the distance between the user and the tablet is small (\circled{1}), Voltron and MP/TP satisfy the QoS constraint. However, as the user moves away from the tablet (i.e., from \circled{1} to \circled{3}), TP violates the latency constraints due to the increased communication overhead under weakened signal strength, whereas Voltron still satisfies the latency constraint through communication scaling while maintaining nearly the same accuracy. When the user moves even farther, the device becomes disconnected from the cluster (\circled{4}). In this case, both MP/TP cannot execute the full LLM due to the reduced memory budget, resulting in severe accuracy degradation. Even in this case, Voltron sacrifices only a small amount of accuracy by exploiting computation scaling --- changing all layers to INT4 and additionally applying pruning, thereby enabling execution under the reduced memory budget.

\begin{figure}[t!]
\includegraphics[width=0.98\linewidth]{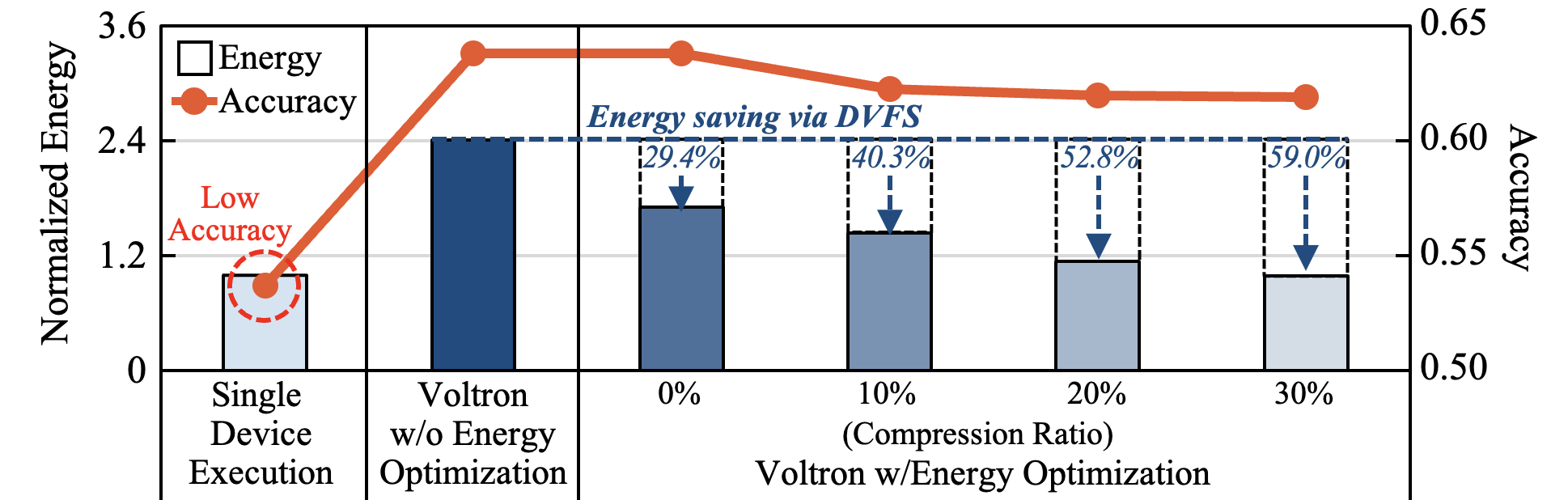}
\caption{Impact of Voltron Energy Optimization}
\label{fig:Evaluation5}
\vspace{-0.3cm}
\end{figure}

\subsubsection{Energy Analysis}
\label{Section5.2.3}
When Voltron adopts the energy optimization with different compression ratio (i.e., how much parameters of a LLM is reduced via precision scaling and pruning), the energy consumption can be significantly reduced via DVFS --- the energy consumption can even be reduced by 59.0\% which is almost the same as that of the single execution baseline (less than 4.4\% of difference). Voltron does not incur a severe accuracy drop even in those cases (only 1.9\% even under the 30\% compression ratio), by adopting the importance-aware principle to the precision scaling and pruning. This result implies that \textit{Voltron can even strike better accuracy-energy trade-off point compared to the single device execution}, making the multi-device LLM execution more promising for future applications.

\subsubsection{Overhead Analysis}
\label{Section5.2.4}
Voltron elastically adjusts the layer-wise execution strategy and precisions only with a negligible overhead.
The execution plan module incurs a negligible overhead (average of 0.1ms in our experiments), which only accounts for 0.03\% of TPOT. The precision allocation module also exhibits a small overhead (average of 0.8ms in our experiments) thanks to the low complexity of binary search --- naive enumeration takes up to 24.1ms in the worst case though. Voltron also does not incur severe overhead for the actual plan modifications. Precision scaling and pruning incur I/O overhead, but Voltron mitigates the I/O overhead 1) by overlapping I/O operations with on-going computations, and 2) with shard-wise memory allocations. This decreases the modification overhead from 31.7ms to 4.1ms per output token. As a result, the total overhead accounts for only 1.2\% of TPOT, making it feasible for Voltron to determine an adequate execution plan token-by-token.

\section{Related Work}

\noindent
\textbf{Distributed Inference Method:} With the growing size of LLMs, there have been increasing pushes to execute LLM inference across multiple compute nodes (e.g., GPUs) in clusters. To run LLM inference on the multiple compute nodes, prior works have explored various distributed inference techniques, such as tensor parallelism~\cite{megatronlm} and model parallelism~\cite{huang2019gpipeefficienttraininggiant,lepikhin2020gshardscalinggiantmodels}, which partition model weights and computations across multiple compute nodes. However, most existing approaches are adopting a single parallelism technique uniformly across all layers of an LLM. Although these techniques are effective in homogeneous data center clusters, they offer limited flexibility in heterogeneous or resource constrained environments~\cite{PerformanceAwareMP, performanceAwareTP,HelixAsplos}. In addition, modern LLM architectures consist of diverse layer types with different computational and memory characteristics~\cite{qwen2025qwen25technicalreport, bai2023qwentechnicalreport, gemmateam2025gemma3technicalreport, abdin2024phi3technicalreporthighly}, making the parallelism methods have different performance characteristics and memory footprint~\cite{huang2019gpipeefficienttraininggiant,lepikhin2020gshardscalinggiantmodels,zheng2022alpaautomatinginterintraoperator}. Different from the above techniques, Voltron enables flexible layer-wise hybrid parallelism that dynamically adapts execution strategies to heterogeneous devices and resource-constrained environments, thereby enabling efficient distributed LLM inference across edge devices.

\noindent
\textbf{On-Device LLM inference:} There have been increasing pushes to execute LLM inference at the edge, due to privacy concerns. To enable on-device LLM inference, several algorithm-level techniques, such as sLLMs~\cite{tinyllama, mobilellm}, pruning~\cite{SixteenHead, LLMPruner}, and quantization~\cite{AWQ, SmoothQuant, GPTQ}, have been proposed. sLLMs reduce model size by designing lightweight architectures tailored for edge environments. sLLMs are typically trained from the scratch~\cite{mobilellm, tinyllama} or distilled from the larger LLMs~\cite{llama3}. Pruning partially removes parameters in a structured~\cite{LIANG2021370} or unstructured~\cite{li2022parameterefficientsparsitylargelanguage} manner, based on their importance~\cite{SixteenHead}. Quantization reduces memory and computation by representing model parameters in low precision. Quantization adopts low precision (e.g., INT4) to entire parameters~\cite{banner2019posttraining4bitquantizationconvolution} or partial parameters in various granularity~\cite{10.1145/3503161.3548001,dettmers2022llmint88bitmatrixmultiplication,10.1145/3575693.3575698}. Though the above techniques effectively reduces resource requirements, they often incur non-trivial accuracy loss due to the reduced model capacity or information loss~\cite{nagel2021whitepaperneuralnetwork}. In contrast, Voltron enables on-device LLM inference through a system-level approaches that flexibly leverages multiple edge devices to execute larger models while minimizing accuracy loss. Several system-level works have tried to enable on-device execution of LLMs, of which parameter size is beyond the single device memory~\cite{xue2024powerinfer2fastlargelanguage,alizadeh2024llm,sheng2023flexgen}, at the expense of the increased latency. Voltron can be complementarily adopted along with such works, to further maximize the accuracy of LLM inference at the edge while satisfying the QoS constraints. 

\section{Conclusion}


To break through memory limitations of on-device LLM inference execution, in this paper, we propose Voltron, a novel on-device LLM inference framework which elastically exploits multiple devices adapting to the execution environments. Our characterization reveals that the performance and accuracy of multi-device LLM inference are strongly affected by characteristics of parallelism methods, diverse edge heterogeneity, and runtime variance. To address the issues, Voltron adopts layer-wise hybrid parallelism and mixed precision allocation, and elastically adjusts them adapting to the execution environment changes. Voltron achieves 10.2\% higher accuracy, on average, compared to the baselines while satisfying QoS constraints. We believe Voltron will empower the edge intelligence for future applications.


\bibliographystyle{ACM-Reference-Format}
\bibliography{refs}

\end{document}